\documentclass[10pt]{article} 
\usepackage{epsfig} 
\usepackage{amsfonts}
\usepackage{graphicx}
\usepackage{epsfig}

\usepackage{a4} 
\usepackage{amsmath}

\date{\today}

\newcommand{\si}{\sigma}

\newcommand{\ee}{\end{equation}}
\newcommand{\eea}{\end{eqnarray}}
\newcommand{\be}{\begin{equation}}
\newcommand{\bea}{\begin{eqnarray}}

\newcommand{\re}[1]{(\ref{#1})}

\begin{document}
 
\title{ Einstein-Yang-Mills solutions in higher
\\ dimensional de Sitter spacetime}
\author{{\large Yves Brihaye,}$^{\ddagger}$ {\large Eugen 
Radu}$^{\dagger}$
and {\large D. H. Tchrakian}$^{\dagger \star}$ 
\\ 
\\
$^{\ddagger}${\small Physique-Math\'ematique, Universite de
Mons-Hainaut, Mons, Belgium}
\\ 
$^{\dagger}${\small Department of
Mathematical Physics, National University of Ireland Maynooth,}
{\small Maynooth, Ireland} 
\\
$^{\star}${\small School of Theoretical Physics -- DIAS, 10 
Burlington
Road, Dublin 4, Ireland }}

\maketitle

\begin{abstract}
We consider particle-like and black holes solutions of the
Einstein-Yang-Mills system with positive cosmological constant in 
$d>4$ spacetime
dimensions. These configurations are spherically symmetric
and present a cosmological horizon for a finite value of the radial 
coordinate,
approaching asymptotically the de Sitter background. In the usual 
Yang--Mills
case we find that the mass of these solutions, evaluated outside the 
cosmological
horizon at future/past infinity generically diverges for $d>4$.
Solutions with finite mass are found by adding
to the action higher order gauge field terms belonging to the 
Yang--Mills
hierarchy. A discussion of the main properties
of these solutions and their differences from those to the usual 
Yang--Mills
model, both in four and higher dimensions is presented.
\end{abstract}

\section{Introduction}
Recently, there has been a significant increase 
in interest in the properties of gravity in more than $d=4$ 
dimensions. 
This interest was enhanced with the development of string theories, 
along with the idea of large extra 
dimensions recently resurrected by TeV gravity models. 
Several solutions of higher dimensional classical general 
relativity have been known for some time, 
in particular extensions to any $d>4$ of the 
Schwarzschild and Reissner-Nordstr\"{o}m black 
holes by Tangherlini \cite{Tangherlini:1963bw}, 
and of the Kerr black hole by Myers and Perry \cite{Myers:1986un}. 
Investigations over the past fifteen years have produced an
impressive catalogue of solutions for various effective
theories of Einstein gravity coupled to many different kinds of 
matter fields.
These results indicate  that the
physics in higher-dimensional general relativity
is far richer and complex than in the standard four-dimensional 
theory.

Solutions to the Einstein-Yang-Mills (EYM) equations in higher 
dimensions have
recently been studied. As found in \cite{Volkov:2001tb}, for 
asymptotically flat
solutions to the usual Yang-Mills (YM) gravitating system in five 
spacetime
dimensions the particle spectrum obtained by uplifting the $d=4$ flat 
space YM
instantons become completely destroyed by gravity, as a result of 
their scaling
behaviour. These results~\cite{Volkov:2001tb} can be systematically 
extended to
the $d\ge 5$ case and one finds that no finite mass spherically 
symmetric 
solutions exist in EYM theory, unless one modifies the 
non Abelian action density by adding higher order curvature terms 
in the YM hierarchy. (The YM hierarchy features higher order 
curvature $2p$
forms $F(2p)=F\wedge F\wedge F...\wedge F$, $p$ times, labeled by $p$, 
the
$p=1$ term giving the usual YM system.)
Without these higher order YM terms, only vortex-type finite energy 
solutions
\cite{Volkov:2001tb} exist, describing effective systems
in three spacelike dimensions and with a number of codimensions. 

Asymptotically flat, regular, static and spherically symmetric 
solutions of EYM
equations with higher order terms in the Yang-Mills hierarchy were
presented in \cite{Brihaye:2002hr} for
spacetime dimensions $d=6,7,8$, and for $d=5$, both
globally regular and black hole solutions were found in 
\cite{Brihaye:2002jg}. 
The properties of these solutions are rather different from the 
familiar
Bartnik-McKinnon solutions~\cite{Bartnik:1988am} to EYM equations in 
$d=4$, and are
somewhat more akin to the gravitating monopole solutions to EYM-Higgs 
system
\cite{Breitenlohner:1992}. This is because like in the latter case
\cite{Breitenlohner:1992}, where the vacuum
expectation value of the Higgs field features as an
additional dimensional constant, here  also additional
dimensional constants enter with each higher order YM curvature
term \cite{Brihaye:2002hr}.  The typical
critical features discovered in \cite{Brihaye:2002hr,Brihaye:2002jg} 
have
been analysed and explained in \cite{Breitenlohner:2005hx}.

When a cosmological constant is added to the theory, the asymptotic
behaviour of the physically relevant solutions
changes from Minkowski to Anti-de Sitter (AdS) for $\Lambda<0$ or de 
Sitter (dS)
for $\Lambda>0$. In the latter case ($\Lambda>0$), to our knowledge 
no studies of
EYM solutions in higher dimensions ($d>4$) have been undertaken to 
date. To carry out
this investigation is the aim for the present work.

Higher dimensional EYM-$F(2)$  solutions with a negative cosmological 
constant ($\Lambda<0$)
have been studied in \cite{Okuyama:2002mh, Radu:2005mj}.
The  main properties of these 
configurations resemble the familiar $d=4$ AdS ones 
\cite{Winstanley:1998sn,Bjoraker:2000qd}.
These describe a continuum of solutions with arbitrary asymptotic 
values of the
function $w(r)$ parametrising the spherically symmetric gauge field.
However, the total mass-energy of the AdS$_d$ nonabelian solutions 
diverges for $d>4$.
Higher dimensional asymptotically AdS EYM solutions with finite 
mass-energy  
were found by augmenting the action density of the system with higher 
order curvature
terms, consisting of $2p$-form curvatures $F(2p)$ \cite{Radu:2005mj}.
The qualitative features of these solutions are very similar to those 
of the
asymptotically flat case with $\Lambda=0$; in particular, and most 
notably, the
nonabelian fields approach asymptotically a pure gauge
configuration with $w(r\to\infty)=-1$, uniquely, unlike in the $p=1$ case.

Proceeding to EYM systems with positive cosmological constant 
$\Lambda>0$, we note that
at present only the $p=1$ gravitating YM system is studied, and that,
only in $d=4$
(see \cite{Volkov:1996qj}-\cite{Torii:1995wv}, and also the 
systematic approach
in \cite{Breitenlohner:2004fp}). The gravitating YM field
equations present both black holes and solutions with a regular 
origin, in contrast
to the case of a gravitating Abelian field. This property is shared 
with all cases,
$\Lambda>0$, $\Lambda<0$ and $\Lambda=0$.
Asymptotically dS ($\Lambda>0$) configurations present, in 
particular, a cosmological
horizon for a finite value of the radial coordinate.
As in the $\Lambda<0$ case,
the asymptotic value of the gauge field function $w(r)$ for solutions 
with $\Lambda>0$
is not fixed, implying the existence of a nonvanishing magnetic 
charge. This
contrasts with the asymptotically flat situation 
\cite{Bartnik:1988am}.

Our task in this paper is to examine the corresponding situation
for $d>4$ EYM solutions with positive cosmological constant.
Our strategy is to first consider the usual YM model, namely the 
$p=1$ member
of the YM hierarchy, i.e. the square of the $2$-form curvature 
$F(2)$.
We find that although the EYM
equations in this case present solutions approaching asymptotically 
the dS
background, the mass of solutions evaluated at future/past infinity 
generically diverges.
Like in the asymptotically flat and AdS cases, finite mass-energy EYM 
solutions are
found by augmenting the action density of the system with higher 
order curvature terms,
consisting of $2p$-form curvatures $F(2p)$.

\section{Higher dimensional gravitating $p=1$ YM system}
\subsection{The model}
In this Section we shall examine the usual EYM system
in a $d-$dimensional spacetime described by the following action
\begin{eqnarray}
\label{action}
I=I_{bulk}+I_{surf}=\int_{\mathcal{M}} d^d x \sqrt{-g}
\left(
 \frac{1}{16 \pi G}(R-2 \Lambda)
 +{\cal L}_m
\right)
-\frac{1}{8 \pi G}\int_{\mathcal{\partial M}^{\pm }} d^{d-1}x\sqrt{h 
}K ,
\end{eqnarray}
where $R$ is the Ricci scalar associated with the
spacetime metric $g_{\mu\nu}$,  $\Lambda=(d-1)(d-2)/(2 \ell^2)$ is 
the
cosmological constant and $G$ is the gravitational constant 
(following \cite{Brihaye:2002hr,Brihaye:2002jg}, we define also 
$\kappa=1/(8\pi G)$).
${\cal \partial M}^{\pm }$ are spatial Euclidean boundaries at 
Euclidean surfaces 
at future/past timelike infinity $\mathcal{I}^{\pm}$
and $\int_{\mathcal{\partial M}^{\pm }}$ indicates the sum of the
integral over the early and late time boundaries. 
The quantities $g_{\mu \nu },h_{\mu \nu } $ and $K$ 
are the bulk spacetime metric, induced boundary metrics and the trace 
of
extrinsic curvatures of the boundaries respectively. 

The matter term in the above relation,
${\cal L}_m=-\frac{1}{4} \tau_1 ~{\rm tr}~\{F_{\mu \nu} F^{\mu 
\nu}\}$
is the usual $F(2)$ nonabelian action density,
$F_{\mu\nu}=\partial_\mu A_\nu-\partial_\nu A_\mu-i[A_\mu, A_\nu]$ 
being the gauge field 
strength tensor.  
 
The field equations are obtained by varying the action (\ref{action}) 
with
respect to the field variables $g_{\mu \nu},A_{\mu}$ 
\begin{eqnarray}
\label{Einstein-eqs}
R_{\mu \nu}-\frac{1}{2}g_{\mu \nu}R +\Lambda g_{\mu \nu}=
8\pi G  T_{\mu \nu},
~~
 D_{\mu}\left(\sqrt{-g}\ F^{\mu\nu}\right)=0,
\end{eqnarray}
where the energy momentum tensor is defined by
\begin{eqnarray}
\label{Tij}
T_{\mu\nu} =
    {\rm tr}~\{F_{\mu\alpha} F_{\nu\beta} g^{\alpha\beta}
   -\frac{1}{4} g_{\mu\nu} F_{\alpha\beta} 
F^{\alpha\beta}\}.
\end{eqnarray}
For the case of a $d$-dimensional spacetime, we restrict to static 
fields that
are spherically symmetric in the $d-1$ spacelike dimensions, with a
metric ansatz in terms of Schwarzschild coordinates
\begin{eqnarray}
\label{metric}
ds^{2}=\frac{dr^2}{N(r)}+r^{2}d \Omega_{d-2}^2-\sigma^2(r)N(r)dt^2,
\end{eqnarray}  
with  $d\Omega_{d-2}$ the $d-2$ dimensional 
angular volume element and
\begin{eqnarray}
\label{N}
N=1-\frac{2m(r)}{\kappa r^{d-3}}-\frac{r^2}{\ell^2},
\end{eqnarray}
the function $m(r)$ being related to the local mass-energy density up 
to some
$d-$dependent factor.

The choice of the gauge group 
compatible with the symmetries of the line element (\ref{metric}) 
is discussed in \cite{Brihaye:2002hr}. 
This choice implies the use of the representation
matrices $SO_{\pm}(\bar d)$, where $\bar d=d$ and $\bar d=d-1$ for
{\it even} and {\it odd} $d$ respectively.
In this unified notation (for odd and even $d$), the spherically 
symmetric
Ansatz for the $SO_{\pm}(\bar d)$-valued gauge fields then reads
\cite{Brihaye:2002hr}
\begin{equation}
\label{YMsph}
A_0=0\ ,\quad
A_i=\left(\frac{1-w(r)}{r}\right)\Sigma_{ij}^{(\pm)}\hat x^j\ , 
\quad
\Sigma_{ij}^{(\pm)}=
-\frac{1}{4}\left(\frac{1\pm\Gamma_{ \bar d+1}}{2}\right)
[\Gamma_i ,\Gamma_j]\ .
\end{equation}
The $\Gamma$'s denote the $\bar d$-dimensional gamma matrices and
$1,~j=1,2,...,d-1$ for both cases; $\hat x^j=x^j/r$, with 
$r^2=x_ix^i$.

Inserting this ansatz into the
action (\ref{action}), the EYM field equations reduce to
\begin{eqnarray}
\label{eqsF2}
\left(r^{d-4}\si Nw'\right)'
-(d-3)r^{d-6}\si(w^2-1)w=0,
~~
m' =\frac{\tau_1}{2}r^{d-4}\left( 
Nw'^2+(d-3)\frac{(w^2-1)^2}{2r^2}\right),
~~
\frac{  \sigma'}{\sigma}  =\frac{ \tau_1 }{\kappa } \frac{w'^2}{r},
\end{eqnarray}
where the prime denotes the derivative with respect to the radial
coordinate $r$ 

The above differential equations have two analytic solutions. For a 
pure gauge field
$w(r)=\pm 1$, one finds $m(r)=M,~~\sigma(r)=1$, $M$ being a 
constant,
which corresponds to Schwarzschild-dS spacetime.
For $w(r)=0$ we find a non Abelian generalisation of the magnetic-
Reissner-Nordstr{\o}m-dS (RNdS) solution with $\sigma(r)=1$ and
\begin{eqnarray}
\label{RN-sol}
m(r)=M_0+\frac{\tau_1}{2}\log (\frac{r}{\ell}) ~~{\rm for} ~~d=5, ~~{\rm and} ~~
~~m(r)=M_0+\frac{\tau_1(d-3)}{4(d-5)}r^{d-5}
~~{\rm for} ~~d\neq 5, 
\end{eqnarray}
$M_0$ being an arbitrary constant.
Some properties of these solutions are discussed in 
\cite{Gibbons:2006wd}.
We can see e.g. that, for a suitable range of ($M_0,~\Lambda$) it 
describes a 
cosmological black hole, the horizons being located at the zeros of 
$N$.
Also, although these solutions are asymptotically dS, 
their total masses/energies defined outside the 
cosmological horizon at future/past infinity ($r \to \infty$), 
diverge.
 
\subsection{Numerical solutions}
We want the generic line element (\ref{metric}) to describe a 
nonsingular,
asymptotically de Sitter spacetime outside a cosmological horizon 
located at $r=r_c>0$.
Here $N(r_c)=0$ is only a coordinate singularity where all 
curvature invariants are finite. 
\newpage
\setlength{\unitlength}{1cm}
\begin{picture}(18,8)
\centering
\put(2,0.0){\epsfig{file=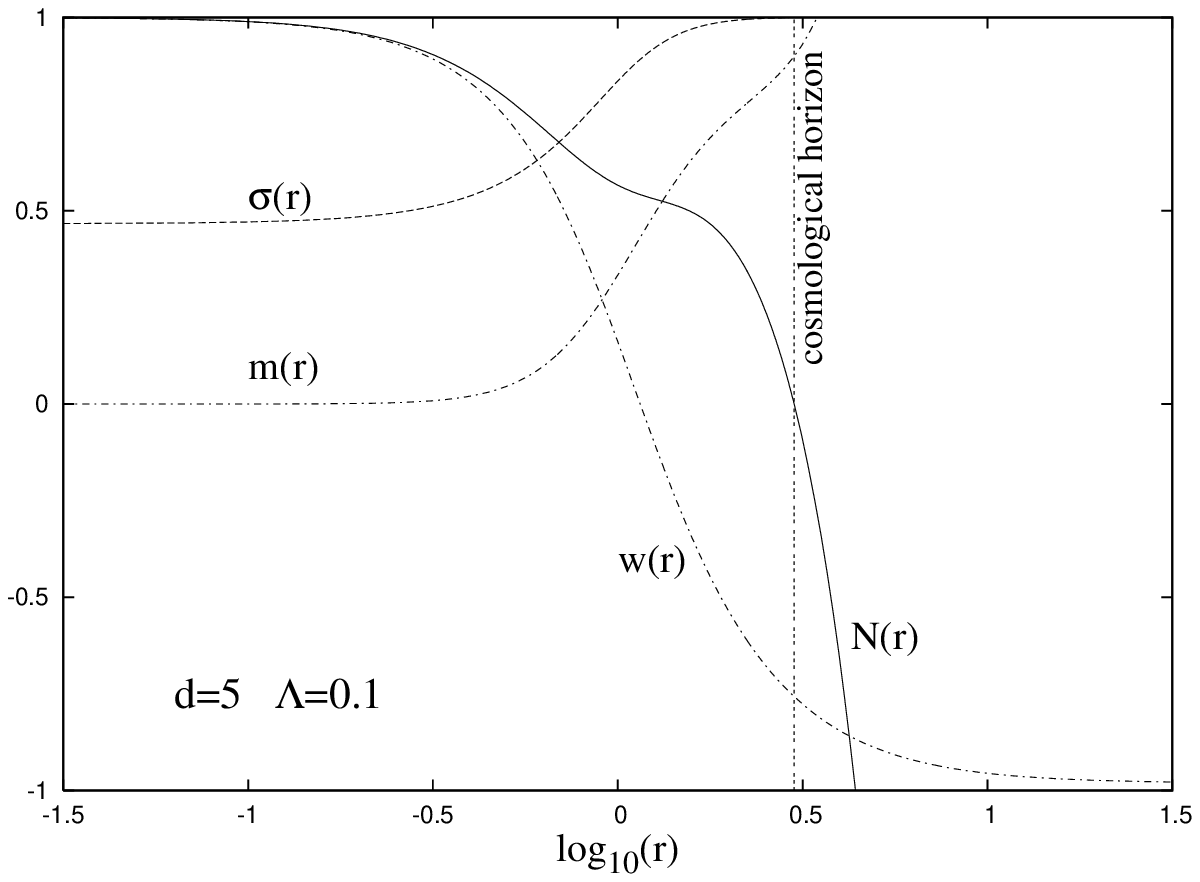,width=11cm}}
\end{picture}
\\
\\
{\small {\bf Figure 1.}
The functions $\sigma(r)$, $w(r)$, $N(r)$ and  $m(r)$
are plotted as functions of radius
for a typical one-node $d=5$ regular solutions in a $F(2)$ EYM-dS 
theory.}
\\
\setlength{\unitlength}{1cm}
\begin{picture}(19,8)
\centering
\put(2.6,0.0){\epsfig{file=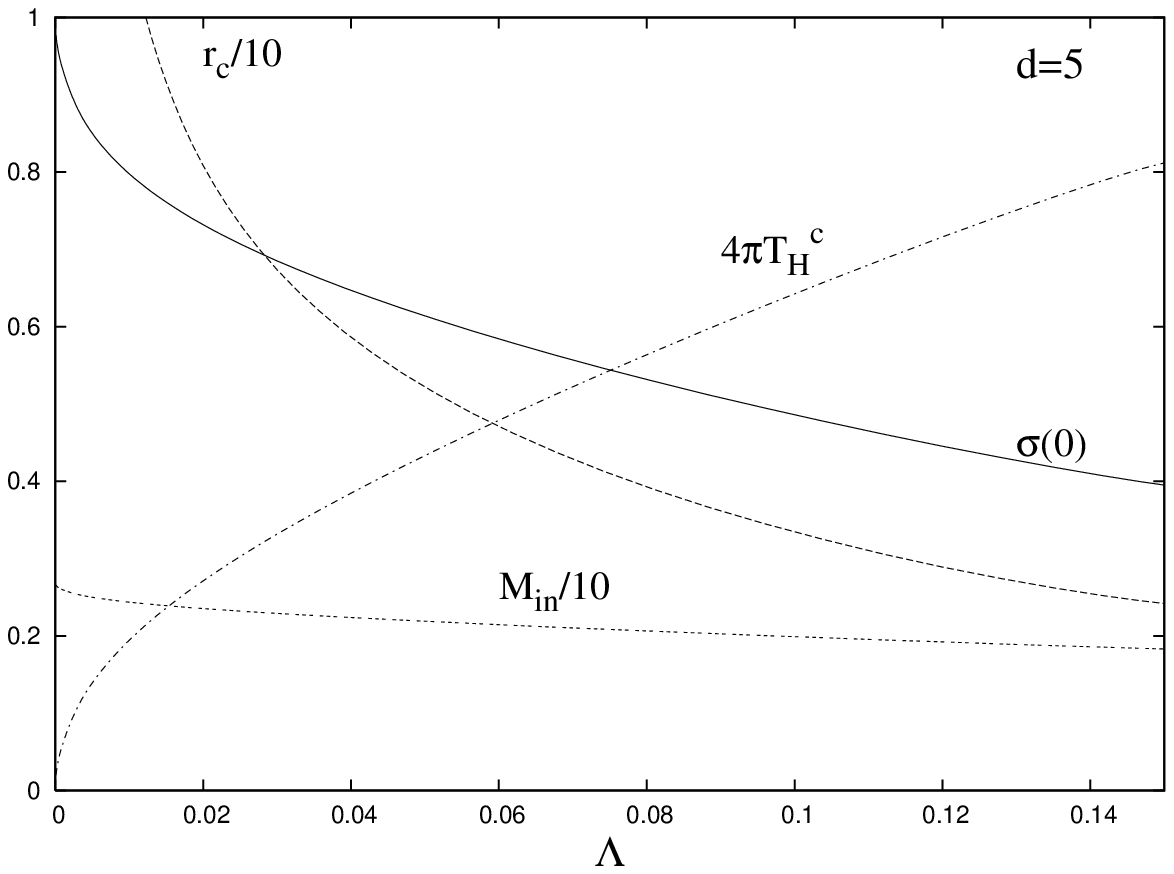,width=11cm}}
\end{picture}
\\
\\
{\small {\bf Figure 2.}
The value of the cosmological horizon radius $r_c$, 
the value $M_{in}$ of the mass function  $m(r)$ at $r=r_c$,
the Hawking temperature associated with the cosmological horizon as
well as the value $\sigma(0)$ of the metric function $\sigma$ at the 
origin,
are shown as functions of $\Lambda$ for $d=5$ 
particle-like solutions of $F(2)$ theory.}
\\
\\
Outside the cosmological horizon $r$ and $t$ changes the character
(i.e.  $r$ becomes a timelike  coordinate  for $r>r_c$).
A nonsingular extension across this null
surface can be found just as at the event horizon of a black hole.
The regularity assumption implies that all
curvature invariants at $r=r_c$ are finite.
Also, all matter functions and their first derivatives extend  
smoothly
through the cosmological horizon, e.g. in a similar way as the U(1) 
electric
potential of a RNdS solution.

\newpage
\setlength{\unitlength}{1cm}
\begin{picture}(18,8)
\centering
\put(2,0.0){\epsfig{file=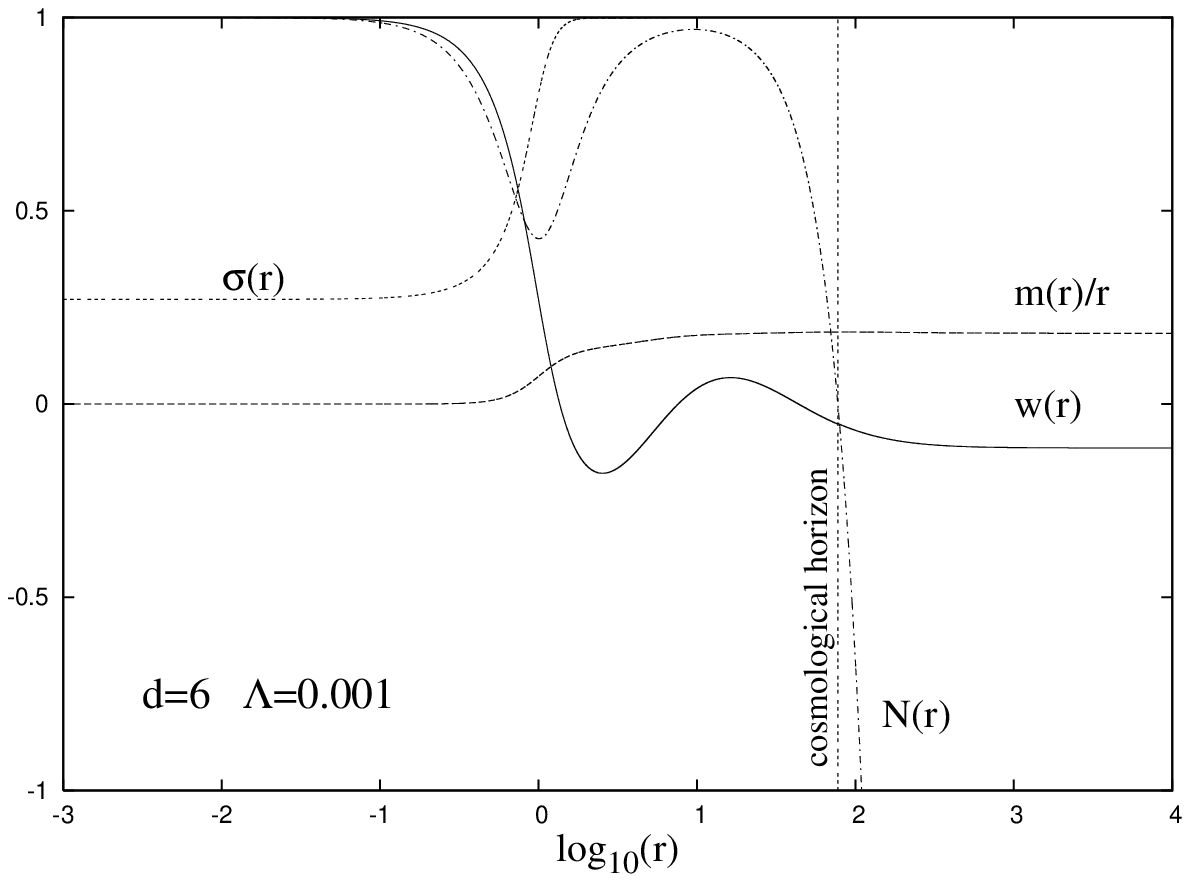,width=11cm}}
\end{picture}
\\
\\
{\small {\bf Figure 3.}
The functions $\sigma(r)$, $w(r)$, $N(r)$ and  $m(r)/r$
are plotted as functions of radius
for a typical  $d=6$ regular solutions in a $F(2)$ EYM-dS theory.}
\\
\setlength{\unitlength}{1cm}
\begin{picture}(19,8)
\centering
\put(2.6,0.0){\epsfig{file=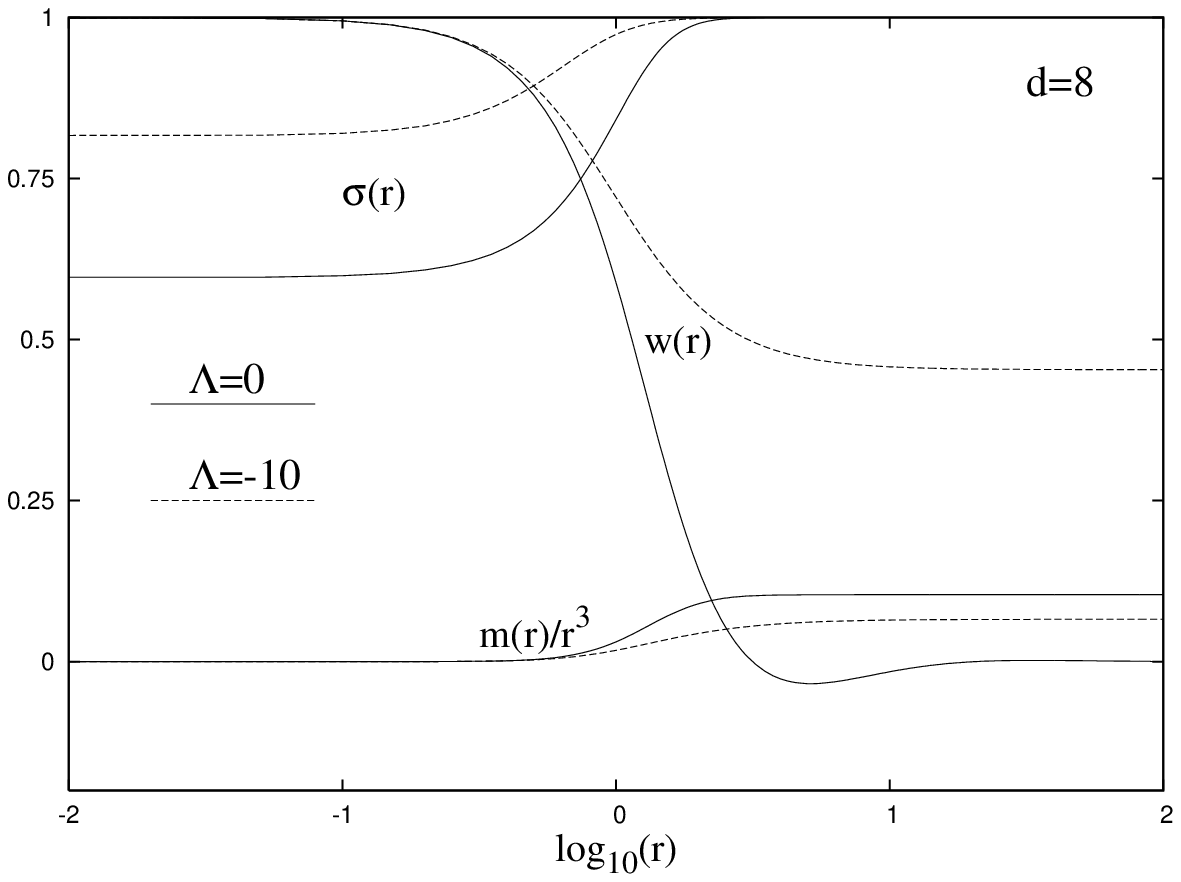,width=11cm}}
\end{picture}
\\
\\
{\small {\bf Figure 4.}
Typical EYM-$F(2)$ solutions in $d=8$ asymptotically flat and  AdS 
spacetimes.}
\\
\\
As in the $\Lambda \leq 0$ case, 
it is natural to consider two types of configurations,
corresponding in the usual terminology to 
cosmological \emph{particle-like} and \emph{black hole} solutions. 
The black hole configurations possess an event horizon located at 
some intermediate value of the radial coordinate $0<r_h<r_c$, all 
curvature
invariants being finite as $r \to r_h$. Both the event and the 
cosmological horizons have their own  surface gravity $\kappa$ 
given by
\begin{eqnarray}
\nonumber
\kappa^2_{h,c}=- \frac{1}{4}g^{tt}g^{rr}(\partial_r 
g_{tt})^2\Big|_{r=r_h,r_c}~~,
\end{eqnarray} 

\setlength{\unitlength}{1cm}
\begin{picture}(19,8)
\centering
\put(2.6,0.0){\epsfig{file=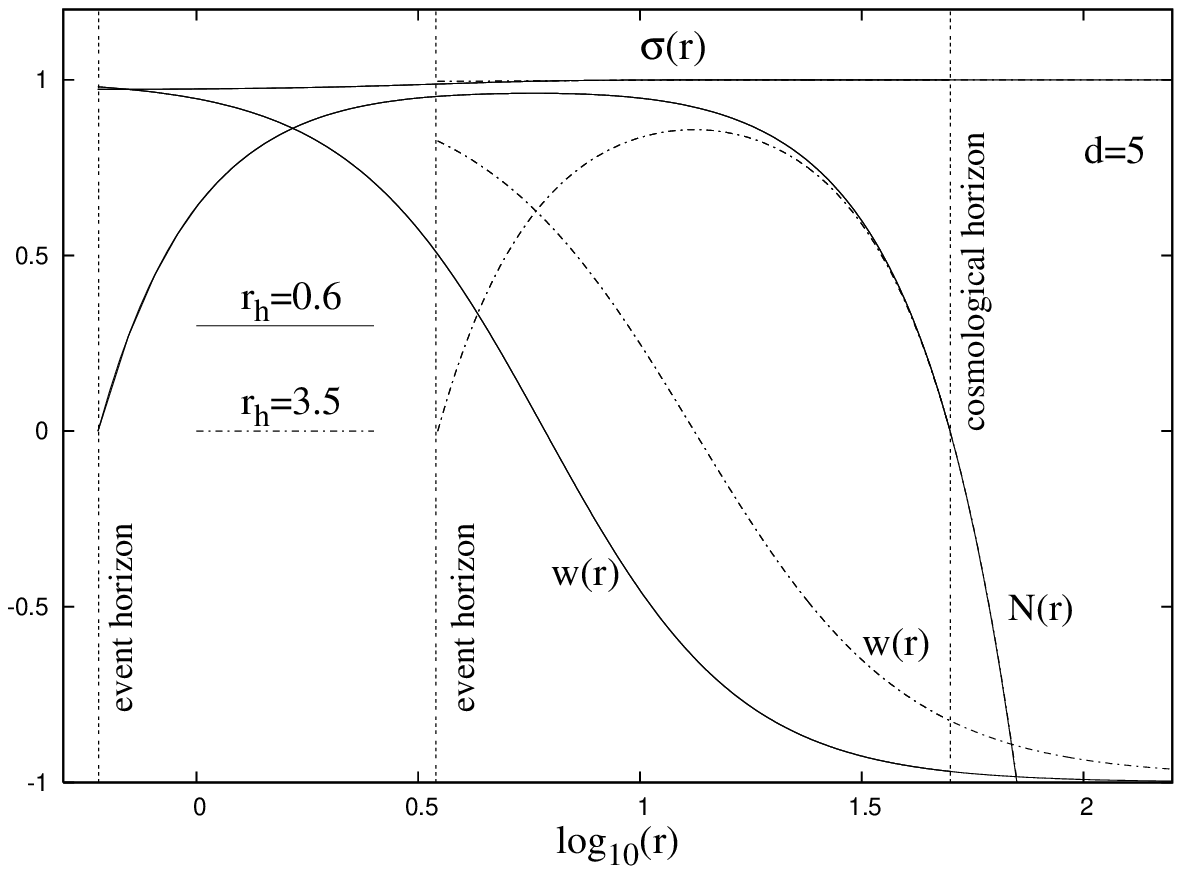,width=11cm}}
\end{picture}
\\
\\
{\small {\bf Figure 5.}
The profiles of two different
$d=5$ EYM-$F(2)$ black hole solution 
are presented for the same value of the cosmological horizon 
radius.}
\\
\\
the associated Hawking temperatures being $T_H^{h,c}
=|\kappa_{h,c}|/(2\pi)$.

The corresponding boundary conditions at the origin and 
cosmological/event
horizon are found by taking $\tau_2=0$
in the general relations (\ref{r=0}), (\ref{eh1})  
given in Section 3.
  
To integrate the field equations, we used in this work the 
differential
equation solver COLSYS which involves a Newton-Raphson method
\cite{COLSYS}.
 The case $\Lambda > 0$ leads to the occurrence of a cosmological
 horizon 
at $r = r_c$ with $N(r_c) = 0$. As in the U(1) case,
the cosmological horizon radius $r_c$ is a function of $\Lambda$. 
In practice, we solved the equations
first on the interval $[0,r_c]$ (or $[r_h,r_c]$), choosing $r_c$ by 
hand and imposing
regularity conditions of the solutions at $r=r_c$.
This allows to determine the values of the functions
$(m,\sigma,w,w')$ at $r=r_c$ as well as the numerical value of 
$\Lambda$ 
corresponding to the choice of cosmological horizon. 
The integration on $[r_c,\infty)$ can then be performed as a second 
step, leading
to the knowledge of the solution on the full $r$-axis and regular at 
$r=r_c$. 

For a $F(2)$ theory, the constants $\kappa$ and $\tau_1$ can always 
be absorbed by
rescaling $r \to cr$, $\Lambda \to \Lambda/c^2$  and $ m \to m \tau_1 
c^{d-5}$,
with $c=\sqrt{\tau_1/\kappa}$, the only remaining parameter
being the cosmological constant. We have numerically integrated Eqs. 
(\ref{eqsF2})
for $d=5,6,7,8$, and for several values of $\Lambda$. 
 
All numerical solutions we found have $w^2(r)\leq 1$,
which implies a nonzero node number of the gauge potential $w$.
This can be proven by using the sum rule 
\begin{eqnarray}
-\left(\frac{r^{d-4}N\sigma w'}{w}\right)\Big|_{r_0}^{r_c}
=\int_{r_0}^{r_c}dr~\sigma \left((d-3)r^{d-6}(1-w^2) +
r^{d-4}N\frac{w'^2}{w^2}\right),
\end{eqnarray}
which follows directly from the YM equations 
($r_0$ here is $r_0=0$ for solutions with regular origin or 
$r_0=r_h$,
for black holes).
Suppose that $w(r)$ never vanishes and $w^2\leq 1$ for $r_0<r<r_c$.
Then the l.h.s. of the above relation vanishes, while the
integrand of the  r.h.s.  is positive definite. Therefore the gauge 
potential
of the nontrivial YM configurations with $w^2\leq 1$ must vanish at 
least once
in the region inside the cosmological horizon.
For solutions with dS asymptotics, the asymptotic value of the
gauge potential $w(\infty)=w_0$ appears as a result of the numerical 
integration.
As a general feature, all configurations we have considered have 
$w_0^2<1$.

Considering first the case $d=5$, 
we were able to construct a numerical solution for each value of the 
cosmological
constant $\Lambda \leq 0.5$, 
the numerics becoming more difficult for larger values of $\Lambda$. 
As in the $d=4$
case \cite{Volkov:1996qj}, we expect there to be a maximal value 
$\Lambda_c$
of the cosmological constant, such that for $\Lambda>\Lambda_c$ the 
gravitational
interaction becomes too strong for the solutions with dS asymptotics 
to exist.
The profile of a typical one-node solution corresponding
to $\Lambda=0.1$ and $r_c=3$ is given on Figure 1
(multi-node solutions have been found as well).
One can see that the gauge field function approaches asymptotically a 
finite value,
$w_0\simeq -0.98$ as a consequence of which the mass function 
diverges logarithmically.

The corresponding numerical value of $r_c, \sigma(0)$, the Hawking 
temperature 
$T_H$ of the cosmological horizon  
and the value of the mass function on the cosmological horizon
$M_{in}=m(r_c)$ are reported on Figure 2 as functions of the 
cosmological
constant for one-node configurations.
 One can see that, similar to the case of a RNdS solution, 
the  Hawking temperature increases with $\Lambda$.

Extending these solutions outside the cosmological horizon reveals, 
however,
that the value $w_0$ differs slightly from one.
As a general feature,  the Einstein equations
imply that the mass function $m(r)$ develops a logarithmic
term which makes the mass divergent as $r \to \infty$,
\begin{eqnarray}
\label{mass-div1}
m(r)=M_0+\frac{\tau_1 }{2}(w_0^2-1)^2\log(\frac{r}{\ell}).
\end{eqnarray}
A similar result holds for $d=5$ asymptotically AdS solutions 
\cite{Radu:2005mj},
the $\Lambda=0$ picture being more complex \cite{Volkov:2001tb}.

The results we found by solving numerically the field equations
for $d=6,7,8$  confirm that this is a generic behaviour
of the higher dimensional EYM solutions.
For $d>5$ the mass function diverges for large $r$ according to
\begin{eqnarray}
\label{mass-div2}
m(r)=M_0+\frac{\tau_1(d-3)}{4(d-5)}(w_0^2-1)^2r^{d-5},
\end{eqnarray}
other properties of these solutions being very similar to the the  
$d=5$ case.
Despite this divergence, these solutions are 
still asymptotically dS (as $r \to \infty$ one finds $\sigma \simeq 
1+O(1/r^6)$ ).
A typical $d=6$ configurations with a regular origin is presented
in Figure 3, for $\Lambda=0.0001$ and $r_c=77.45$.
One can see that the mass function diverges linearly, while
the gauge potential $w(r)$  presents three nodes.

For completeness, we give in Figure 4 the profiles of 
typical EYM-F(2)  $d=8$ configurations in asymptotically flat and AdS spacetimes
(the solutions for other dimensions $d>5$ have the same qualitative features).
The solutions we have found   for $\Lambda \leq 0$
exist for a compact interval $0<b<b_{max}$,
$b$ being the parameter in the expansion of the gauge function at the origin
$w(r)=1-br^2+O(r^4)$, their masses diverging again according to (\ref{mass-div2}),
since $w_0^2\neq 1$. For the AdS case, the asymptotic value of the gauge potential 
may take arbitrary values \cite{Radu:2005mj}, being fixed by  the
parameter $b$ (with $b=0.5$ for the solutions in Figure 4).
All asymptotically flat solutions we have studied have $w_0=0$ and present
at least one node of the gauge function $w(r)$.
Also, a critical solution is approached 
as $b\to b_{max}$, with $\sigma(0)\to 0$ in this limit.

Apart from particle-like solutions, we have found black hole 
solutions as well.
Restricting to the $d=5$ case, our numerical analysis suggests 
that any 
asymptotically dS particle-like solution presents black hole counterparts.
Imposing the condition of a regular horizon at $r=r_h$, we obtained a 
family
of black hole solutions for any $r_c>r_h >0$. When the value $r_h$ 
increases,
the values $|w(r_h)|$ and $|w(r_c)|$ slowly decrease.
The asymptotics of the cosmological black hole solutions are similar 
to
the particle-like case. In particular one finds $w_0^2<1$, which 
implies a divergent
value of the mass-function as $r \to \infty$ according to 
(\ref{mass-div1}).
The profiles of the solutions corresponding two different values
of $r_h$ are reported on Figure 5 (these configurations have the 
cosmological
horizon at $r_c=50$).

We see that the mass function of both regular and black hole 
solutions diverges
asymptotically, yielding infinite total mass. The situation is 
exactly the same for the
$\Lambda\le 0$ too. Not having a finite value for the mass in the 
$\Lambda>0$ case
however may not be regarded as quite as serious a physical 
disadvantage, if one takes
the view that the mass is nonetheless finite inside the cosmological 
horizon.

For $\Lambda\leq 0$, the non-existence of $d>4$ spherically 
symmetric
EYM-$F(2)$ configurations with finite mass could be proven in a 
rigurous way.
However, the arguments in \cite{Okuyama:2002mh,Radu:2005mj}
fail to apply for a positive cosmological constant, a different 
approach 
being necessary in this case.

\section{Higher dimensional gravitating nonabelian $p$ YM 
hierarchies}
\subsection{The equations and boundary conditions}
A simple way to find 
nontrivial solutions with a finite mass is to modify the matter 
Lagrangean
by adding higher order terms in the YM hierarchy,
constructed exclusively from YM curvature $2p$-forms.
Such terms are also predicted by the low energy string theory
(see e.g. \cite{Tseytlin}-\cite{CNT}). 

The definition we use for superposed YM hierarchy is
\be
\label{YMhier}
{\cal L}_{m}=-\sum_{p=1}^{P}\ \frac{1}{2(2p)!}\ \tau_p~\sqrt{-g}~
{\mbox {\rm Tr}\ }\{F(2p)^2\}\ ,
\ee
where $F(2p)$ is the $2p$-form $p$-fold totally antisymmetrised 
product
of the $SO(d)$ YM curvature $2$-form $F(2)$
\be
\label{2pformYM}
F(2p)\equiv 
F_{\mu_1\mu_2...\mu_{2p}}=F_{[\mu_1\mu_2}F_{\mu_3\mu_4}...
F_{\mu_{2p-1}\mu_{2p}]}\ .
\ee
Even though the $2p$-form \re{2pformYM} is dual to a total 
divergence,
namely the divergence of the corresponding Chern-Simons form, the 
density
\re{YMhier} is never a total divergence since it is the square of 
one.
But the $2p$-form \re{2pformYM} vanishes by (anti)symmetry for $d<2p$ 
so
that the upper limit in the summation in \re{YMhier} is 
$P=\frac{d}{2}$
for even $d$ and $P=\frac{d-1}{2}$ for odd $d$.

We define the $p$-stress tensor pertaining to each term in 
(\ref{YMhier}) as
\be
T_{\mu\nu}^{(p)}=
\mbox{Tr}\{ F(2p)_{\mu\lambda_1\lambda_2...\lambda_{2p-1}}
F(2p)_{\nu}{}^{\lambda_1\lambda_2...\lambda_{2p-1}}
-\frac{1}{4p}g_{\mu\nu}\ F(2p)_{\lambda_1\lambda_2...\lambda_{2p}}
F(2p)^{\lambda_1\lambda_2...\lambda_{2p}}\} .
\label{pstress}
\ee 
We shall restrict in this work to solutions
in dimensions less than nine, in which case
it is sufficient to consider the first two terms 
in the YM hierarchy, i.e. a $F(2)+F(4)$ model
(see \cite{ Breitenlohner:2005hx, Radu:2005mj} for the equations 
of the general $(P,d)$ model). As in the previous section, we 
restrict our attention
to static spherically solutions given by the Ans\"atze \re{metric} 
and \re{YMsph},
with exactly the same choices for the gauge group as in Section {\bf 
2.1}.
 
The field equations of the $F(2)+F(4)$ model are
\begin{eqnarray}
 \tau_1\left(\left(r^{d-4}\sigma Nw'\right)'
-(d-3)r^{d-6}\sigma(w^2-1)w\right)+\nonumber \\
+ \frac{\tau_2}{6}(d-3)(d-4)(w^2-1)\left(
\left(r^{d-8}\sigma N(w^2-1)w'\right)'
-(d-5)r^{d-10}\sigma(w^2-1)^2w\right)=0,
\\
m'=\frac12r^{d-4}\Bigg(\tau_1\left[Nw'^2
+\frac12(d-3)\left(\frac{w^2-1}{r}\right)^2\right] 
\nonumber 
\\
+\frac16\frac{\tau_2}{r^2}(d-3)(d-4)\left(\frac{w^2-1}{r}\right)^2
\left[Nw'^2+\frac14(d-5)\left(\frac{w^2-1}{r}\right)^2\right]\Bigg),
\label{meq}
\\
 \kappa \left(
\frac{\sigma'}{\sigma}\right)=\frac{1}{r}
\left[\tau_1+\frac16\frac{\tau_2}{r^2}(d-3)(d-4)
\left(\frac{w^2-1}{r}\right)^2\right]w'^2\ .
\end{eqnarray}
The corresponding expansion of the gauge
potential and metric functions  as $r \to 0$ is
\begin{eqnarray}
\label{r=0}
w(r)&=&1-b r^2+O(r^4),
~~~
m(r)= \Big( \tau_1+\frac{\tau_2}{3}(d-3)(d-4)b^2\Big)b^2 
r^{d-1}+O(r^{d+1}),
\\
\nonumber
\sigma(r)&=&\sigma_0+\frac{2b^2\sigma_0}{\kappa}
\Big(\tau_1+\frac{2\tau_2}{3}(d-3)(d-4)b^2 \Big)
r^{2}+O(r^4),
\end{eqnarray}
and contains one essential parameter $b$ (the value of $ \si_0$
can be fixed by rescaling the time coordinate).

Assuming the existence of a regular, nonextremal event horizon at
$r=r_0$ (with $r_0=r_h$ or  $r_0=r_c$),
the approximate expression of the solution near the event horizon is 

\begin{eqnarray}
\label{eh1}
m(r)=(1-\frac{r_0^2}{\ell^2})\frac{\kappa}{2}r_0^{d-3}+m'(r_0)
(r-r_0)+O(r-r_0)^2,
\\
\nonumber
\sigma (r)=\bar\sigma_0+\sigma_0'(r-r_0)+O(r-r_0)^2,
~~~
w(r)=w_0+w'(r_0)
(r-r_0)+O(r-r_0)^2,
\end{eqnarray}
where
\begin{eqnarray}
\nonumber
m'(r_0)= \frac{r_0^{d-6}}{2}(d-3)(w_0^2-1)^2
\Big(\tau_1+ \tau_2  (d-4)(d-5)\frac{(w_0^2-1)^2}{24r_0^4} \Big),
~~
N'_0=\frac{d-3}{r_0}-\frac{(d-1)r_0}{\ell^2}-\frac{2m'(r_0)}{\kappa 
r_0^{d-3}},
\\
\nonumber
\sigma_0'=
 \frac{\bar\sigma_0 w'^2_0 }{\kappa r_0}
\Big(\tau_1+ \tau_2  (d-3)(d-4)\frac{(w_0^2-1)^2}{6r_0^4} \Big),
~~
w'_0=\frac{1}{N'_0}\frac{w_0(w_0^2-1)}{r_0^2}(d-3) 
(\frac{\tau_1+ \tau_2  (d-4)(d-5)\frac{(w_0^2-1)^2}{ r_0^4} }
{\tau_1+ \tau_2  (d-3)(d-4)\frac{(w_0^2-1)^2}{ 6r_0^4} }),
\end{eqnarray}
with two free parameters, $w_0$, $\bar\sigma_0$.
Also, since the field equations are invariant under
$w \to -w$, one can take $w(0)=1$ and $w(r_h)>0$ without any loss
of generality.

For $r \to \infty$ we find for both regular and black hole solutions
\begin{eqnarray}
\label{29}
w(r)=\pm 1+\frac{w_1}{r^{d-3}}+\dots,~~
m(r)=M-\frac{\tau_1(d-3)w_1^2}{8 \ell^2}\frac{1}{r^{d-3}}+\dots,~~
\sigma(r)=&1-\frac{w_1^2 (d-3)^2\tau_1}{2\kappa 
(d-2)}\frac{1}{r^{d-4}}+\dots.
\nonumber
\end{eqnarray}
These boundary conditions are also shared by the asymptotically flat 
solutions 
(with a different decay of the mass function $m(r)$, however),
$w=\pm 1$ being again the only allowed values of the gauge function 
as
$r \to \infty$. 
As a general feature, 
all solutions discussed in the rest of this section present only one 
node in the gauge function $w(r)$, i.e. $w(\infty)=-1$. 
As in the $\Lambda\leq 0$ cases, we could not find 
multi-node solutions in the $F(2)+F(4)$ model.

\subsection{Numerical solutions}
 
In the presence of higher oder terms in the YM action,
dimensionless quantities are obtained by rescaling 
\begin{eqnarray}
 r \to  (|\tau_2/\tau_1|)^{1/4}r,~~\Lambda \to (|\tau_1/\tau_2|)^{1/2}\Lambda,
~~m(r) \to \kappa(|\tau_1/\tau_2|)^{(d-3)/4}m(r).
\end{eqnarray}
This reveals the existence of one fundamental parameter which gives 
the strength of
the gravitational interaction
$\alpha^2= |\tau_1|^{3/2}/(\kappa |\tau_2|^{1/2})$.
The solutions can then be constructed in terms of $\alpha^2$ and 
$\Lambda$.
Most of the numerical work was carried out for positive
coupling constants in the  YM hierarchy (\ref{YMhier}), having
set  $\tau_1=\tau_2=1$ without loss of generality\footnote{ 
The coupling constant $\tau_1$ equals the inverse of the
square of the gauge coupling constant and
is strictly positive. Also, string theory predicts a
positive value for the coefficient of the $F(4)$
term in the YM hierarchy (see \cite{Tseytlin}-\cite{CNT}).}. After
presenting our main results however, we briefly consider at the end of
this section, the case where $\tau_2$ is taken to be negative. 

Starting again with particle-like solutions in $d=5$, we have solved
equations (\ref{meq}) for several values of $\Lambda$, varying 
$\alpha$. 
The pattern of these solutions  is  illustrated by Figure 6a,
where  the  quantity $\sigma(0)$ is reported
as a function of $\alpha^2$ for three different values of $\Lambda$.

Here it is useful to recall the situation
corresponding to $\Lambda=0$ (see \cite{Brihaye:2002jg}).
In this case, several branches of solutions exist,
depending on the parameter $\alpha^2$, as illustrated on Figure 6a.
The first (or main) branch exists for 
$\alpha^2 \in [0,0.2824]$. Then another branch of solutions (with 
larger masses
than the corresponding one on the main branch) exists for $\alpha^2 
\in [0.1749,0.2824]$.
Several other branches of solutions further exist on 
smaller 
intervals centered on the
critical value $\alpha_{\rm cr}^2 \sim 0.1749$. It is clear from 
Figure 6a that this
oscillatory behaviour converging on $\alpha_{\rm cr}^2$ is a common 
feature of
$\Lambda\ge 0$ solutions. (This is also the case with $\Lambda<0$
solutions~\cite{Radu:2005mj}, which is not displayed on Figure 6a 
since its
branch structure is nearly identical to that of $\Lambda=0$.) 
This phenomenon, which occurs in appropriately similar models in all 
$4p+1$ dimensions,
was exhaustively analysed for the $\Lambda=0$ case 
in~\cite{Breitenlohner:2005hx},
where this $\alpha_{\rm cr}^2$ was called a {\it conical} critical 
point. In the present
paper, we limit ourselves to a qualitative discussion only.

For positive values of the cosmological constant,
our numerical analysis  in the case $0 < \Lambda < 0.001$ reveals 
that
an extra branch of solutions exists, as shown on Figure 6a for 
$\Lambda = 0.0004$. 

\newpage
\setlength{\unitlength}{1cm}
\begin{picture}(19,8)
\centering
\put(2,0.0){\epsfig{file=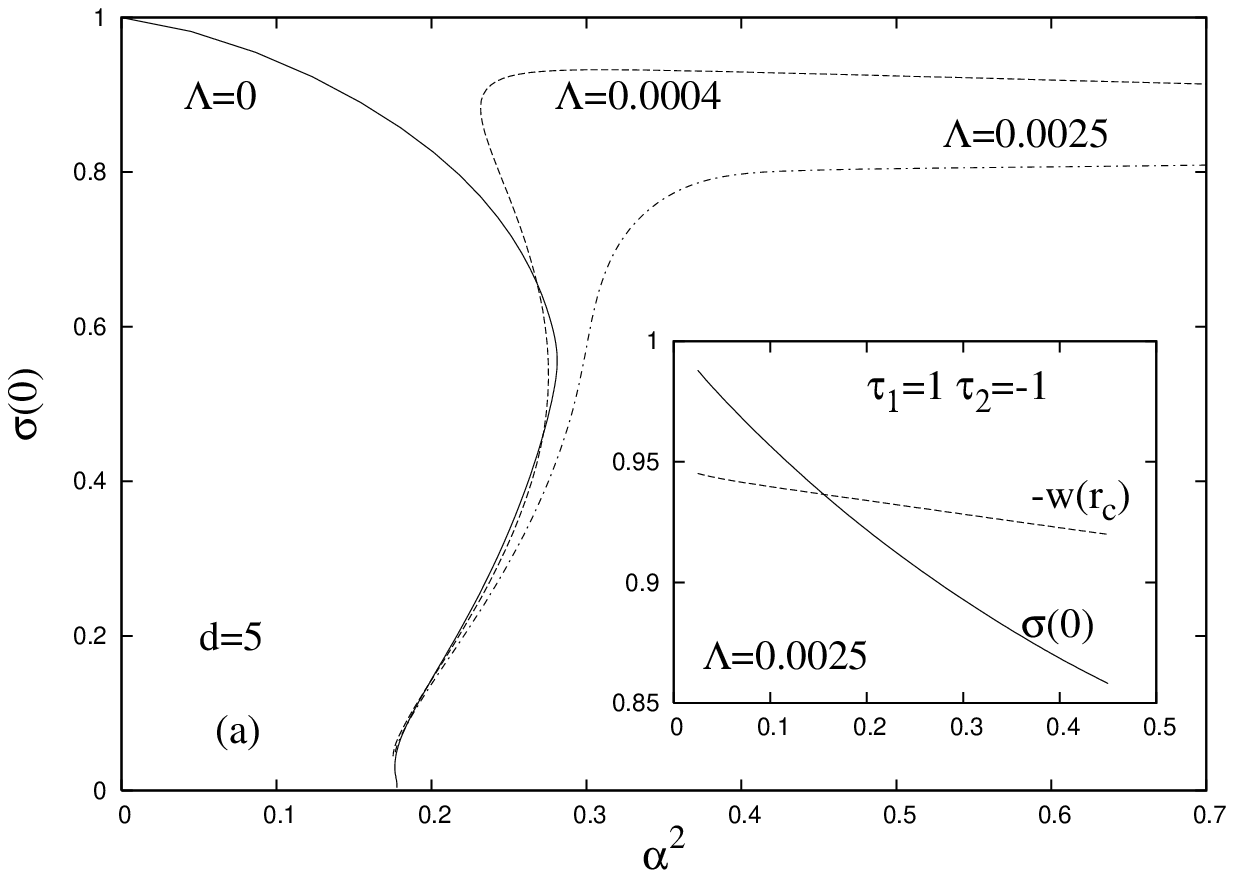,width=11cm}}
\end{picture}
\\
\setlength{\unitlength}{1cm}
\begin{picture}(19,8)
\centering
\put(2.6,0.0){\epsfig{file=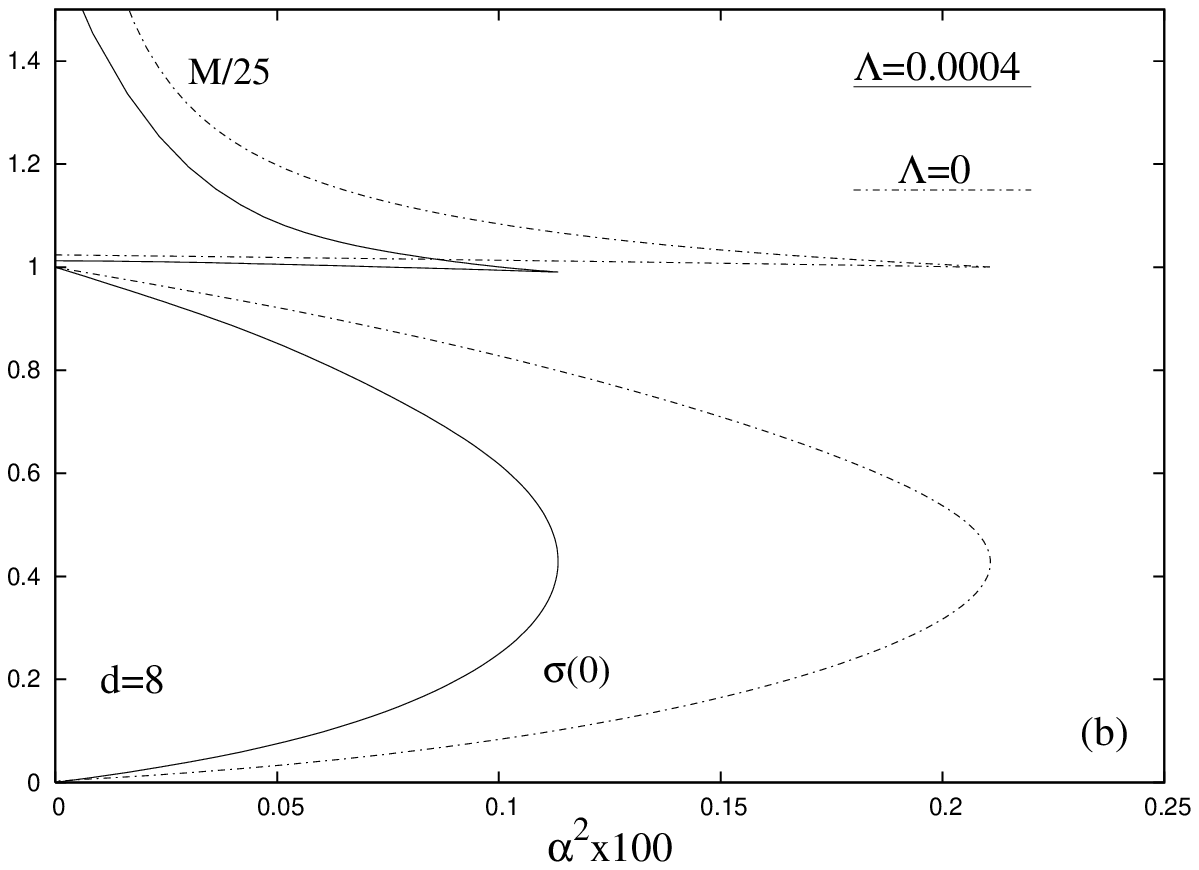,width=11cm}}
\end{picture}
\\
\\
{\small {\bf Figure 6.}
Several global quantities
are shown as functions of $\alpha^2$  and several values of 
$\Lambda$
for $d=5,~8$ solutions of $F(2)+F(4)$ theory.}
\\
\\
Our numerical analysis suggests that this branch,
 which has no counterpart in the $\Lambda = 0$ case, exists for 
$\alpha >\alpha_0^2$, with $\alpha_0^2 \approx 0.21$ for $\Lambda = 
0.0004$.
This branch, appears to survive for arbitrarly large values of 
$\alpha$ (although
the numerical accuracy deteriorates for $\alpha$ larger than one).
Decreasing the cosmological constant, the critical value $\alpha_0^2$ 
decreases and the pattern of the $\Lambda=0$ case is approached. 
For larger values of $\Lambda$ the pattern simplifies and a single 
branch persists, as illustrated for $\Lambda = 0.0025$ on Figure 6a, for $\alpha^2 
>0.1749$.
Also, only a small variation of the asymptotic value $M$ of the mass
function was noticed when varying the parameter $\alpha$.

Three remarks can be made to summarise this description~: (i) For 
$d=5$,
$\Lambda >0$ solutions do not exist for arbitrarily small values of 
$\alpha$;
(ii) Non trivial cosmological solutions seem to exist for large 
values of $\alpha^2$, and
(iii) The multiple branche phenomenon converging on the {\it conical} 
fixed point
$\alpha^2 \sim 0.1749$, observed for 
\newpage
\setlength{\unitlength}{1cm}
\begin{picture}(19,8)
\centering
\put(2.,0.0){\epsfig{file=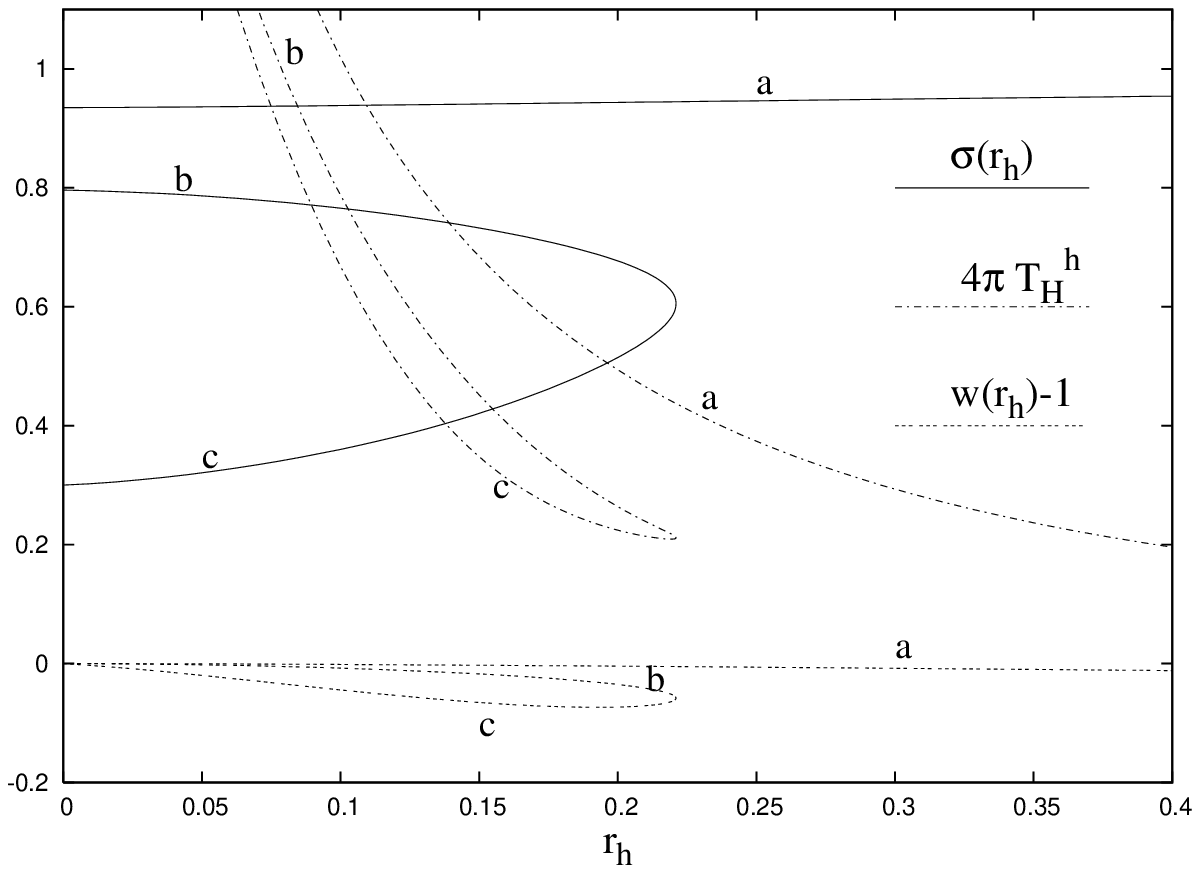,width=11cm}}
\end{picture}
\\
\\
{\small {\bf Figure 7.}
The event horizon Hawking temperature,
the value of the gauge potential on the event horizon $w_h$ as
well as the value of the metric function $\sigma$ at the event 
horizon,
are shown as functions of the event horizon radius for $d=5$ 
black hole solutions of $F(2)+F(4)$ theory with $\alpha^2=0.25$, 
$\Lambda=0.0004$.}
\\
\setlength{\unitlength}{1cm}
\begin{picture}(19,8)
\centering
\put(2.6,0.0){\epsfig{file=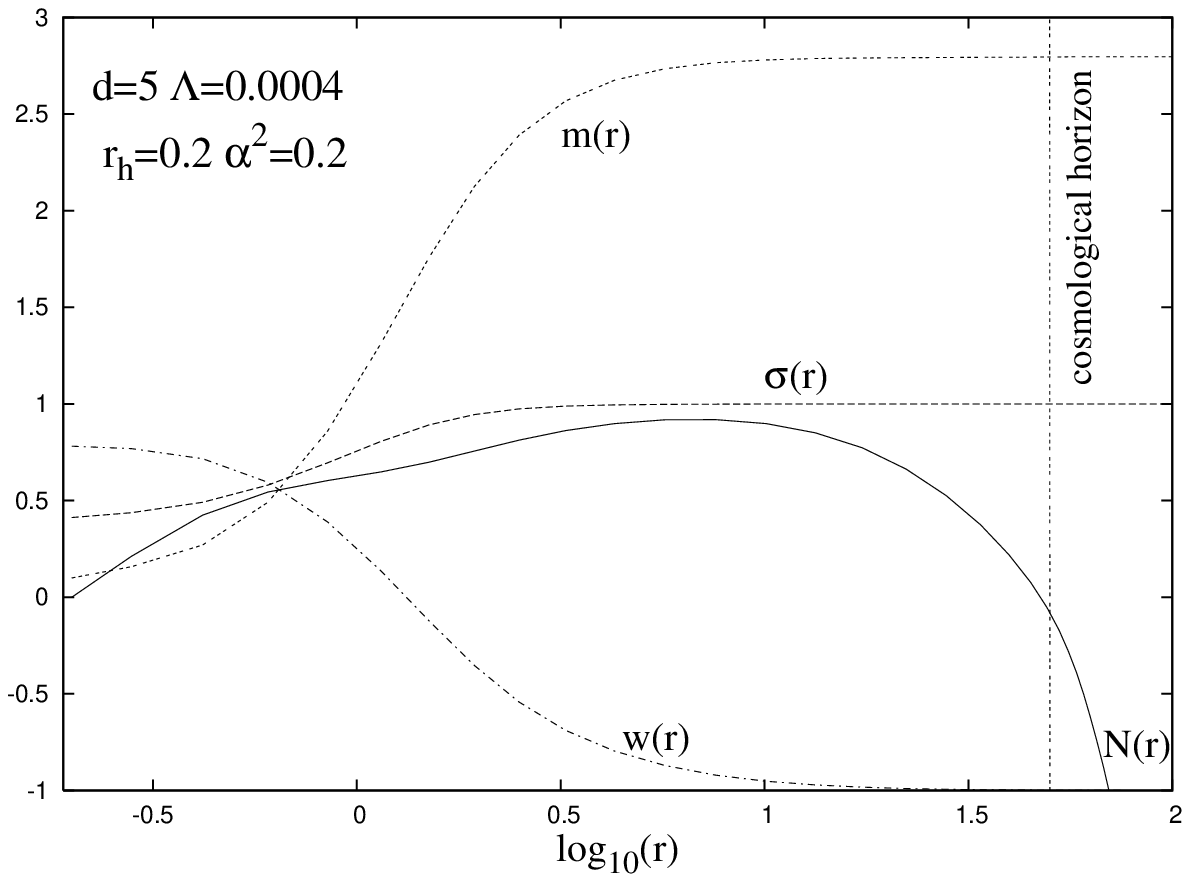,width=11cm}}
\end{picture}
\\
\\
{\small {\bf Figure 8.}
The functions $\sigma(r)$, $w(r)$, $N(r)$ and  $m(r)$
are plotted as functions of radius
for a typical  $d=5$ black hole solution in a $F(2)+F(4)$ EYM-dS 
theory.}
\\
\\
$\Lambda=0$, seems to persist 
for $\Lambda >0$. 

The $F(4)$ term allows also for the existence of $d=6,7,8$ 
configurations with finite
mass. The solutions available in the absence of a cosmological 
constant ($\Lambda = 0$)
feature two branches which exist for $\alpha^2 \in [0,\alpha_{m}^2]$,
where the
maximal value $\alpha_{m}$ depends on $d$ (see 
\cite{Brihaye:2002hr}).
Integrating the equations for $\Lambda >0$ reveals 
that the cosmological solutions also feature two branches leading to 
a pattern very similar to the $\Lambda = 0$ case. 
The maximal values 
$\alpha_{m}$
decrease  with increasing cosmological constant.
Also, with increasing $\alpha$ the  masses of the gravitating 
solutions decrease.
Along the  second branch the value of
$\sigma(0)$ decreases monotonically with $\alpha$. The mass of a 
second branch solution 
is always larger than the corresponding mass (for the same value of
$\alpha$) on the first branch.
Some relevant quantities for $d=8$ solutions are plotted in Figure 
6b.

So far we discussed solutions which are regular at the origin.
However the equations also admit black hole solutions with a regular  
event horizon
occuring at $r_h$ with $r_h < r_c$.  The study of the domain of black 
hole
solutions in the space of the  parameters ($\alpha, \Lambda, r_h$) is 
a considerable
task which is beyond the scope of this paper. We present however
a few features of these solutions which reflect the general pattern, 
limiting
our investigation to the $d=5$ case.
 
As we have seen in the previous section, several branches of regular
solutions exist, according to the value of $\Lambda$. 
When imposing a regular horizon at $r = r_h$,
our numerical analysis indicates that the regular solutions are 
deformed
into black hole solutions. We have analysed in detail the evolution 
of
the black holes solutions in the case $\alpha^2 = 0.25,
\Lambda = 0.0004$ (corresponding to $r_c \simeq 50.0$). In this case
there are three different regular solutions (see Figure 6) 
distinguished namely by the
value of the metric function  $\sigma$ at the origin
(for instance $\sigma(0) \approx 0.93,0.79,0.30$, let us call them
$a$, $b$ and $c$ respectively).
The evolution of these solutions into black holes is summarized on 
Figure 7.
We see clearly that the solutions $b$ and $c$ are deformed into 
black
holes up to a rather small value of $r_h$, namely  for $r_h < 
0.227$.
In the limit $r_h \to 0.227$, the two branches merge into a single 
solution. 
A typical profile of a black hole solution corresponding
to the solution $c$ is presented on Figure 8.

The scenario is completely different for the solution $a$. Indeed, it 
seems that
it can be deformed into a black hole with large event horizon.
When the value $r_h$  increases, we observe that the corresponding 
function
$w(r)$ has a tendency to spead over the interval $r\in [r_h,r_c]$ and 
that
the combination of $w(r),w'(r)$ appearing in the right hand side of
the equation for $m'(r)$ becomes uniformly very small.
As a consequence, the equation for $m(r)$ can be simplified and
leads to the following approximate solution
\be
\label{black}
N(r) \simeq \frac{1}{r^2}
\left(r^2 - r_h^2 - \frac{r^2 + r_h^2}{r_c^2 + r_h^2}(r^2 - 
r_h^2)\right)
\ \ , \ \ \sigma(r) \simeq 1 \ \ , \ \ w(r) \simeq \pm1+\frac{w_1}{r^2}~,
\ee
for the region between the event and cosmological horizons.
The metric functions $N(r)$, $\sigma(r)$ above
turn out to be a very good approximations of the numerical solution 
$a$
obtained for $r_h > 1$ (although $w(r)$ is non constant).
Due to the presence of this solution, the
numerical integration of the equation with the non trivial $w(r)$
becomes increasingly difficult while increasing $r_h$.

The Hawking temperature associated with the cosmological horizon
 is almost constant on the branch $(a)$, and stronly decreases with 
$r_h$ for the other two branches.
As a generic feature, the Hawking temperatures of the event and 
cosmological
horizons are different (this holds also for the black hole solutions 
of the
$F(2)$-theory).
Therefore, the energy flows from the hotter horizon
to the cooler one and the black hole will gain or lose mass.

 Before concluding this section, we allude to the unusual situation
where the coupling constant $\tau_2$ associated with the
$F(4)$ term is taken to be negative~\footnote{Ascribing a negative
value to the square of the gauge coupling constant has been considered
in the case of gravitating electromagnetism in \cite{Gibbons:1996pd},
where explicit solutions are found. We found however that in the
corresponding situation when the Abelian field in \cite{Gibbons:1996pd} is
replaced by a non-Abelian $SU(2)$ gauge field, no numerical
solutions with right asymptotics appear to exist.}.
 We have managed to
construct such solutions, which satisfy the same set of 
boundary conditions as the usual ones above with $\tau_2>0$.
Physically, this would imply a negative contribution
to the mass-energy density of the solutions, but in practice this
possibility does not seem to obtain and the solutions we constructed
lead to finite masses.

Since we do not have any existence proof for the type of EYM solutions
found in this section, we are unable to exclude the possibility of
solutions with $\tau_2<0$. For this reason we have extended our
numerical analysis to $\tau_2<0$ models with $\Lambda \leq 0$ as well as
the $\Lambda > 0$ model at hand.
Our numerical results indicate the absence of such configurations approaching 
asymptotically the AdS or flat background in all dimensions between five and
eight. Surprisingly enough, we found finite mass solutions
with $\tau_2<0$ in the dS case studied here.
The results for $d=5$ indicate the existence  of
a branch of solutions emerging at $\alpha=0$ from a
configuration in fixed dS background and extending in $\alpha$
up to a maximal value $\alpha_{max}$.
This behaviour strongly constrasts with what we have found for $\tau_2>0$.
However, the profile of a typical $F(2)+F(4)$ solution
does not depend on the sign of $\tau_2$.
As $\alpha \to \alpha_{max}$ the numerics deteriorates
and the solver fails to converge
(although all metric and matter functions stay finite there), a different approach being necessary.
Along this branch, the value at the origin
of the metric function $\sigma$ decreases and also the mass-parameter $M$
(note that all solutions we found with $\tau_2<0$ have $M>0$, with a small variation
when increasing $\alpha$, however).
The corresponding picture for $\tau=1,~\tau_2=-1$ and $\Lambda=0.0075$ is plotted
in the inlet of Figure 6a.

\section{A computation of the mass in the boundary counterterm 
method}
In evaluating expressions like the mass and action, one usually 
encounters
divergencies coming from integration over the infinite volume
of spacetime.
In the case of AdS gravity, a regularization 
procedure was proposed in \cite{Balasubramanian:1999re}, that 
consists in adding to (\ref{action}) counterterms constructed from
local
curvature invariants of the boundary.
These counterterms, which are esentially unique, can be easily 
generalized to the
case of a positive $\Lambda$ \cite{Ghezelbash:2001vs}.
The following counterterms are sufficient to cancel divergences 
in  a pure dS gravity theory for 
$d\leq 7$  
\begin{eqnarray}
\label{ct}
I_{\rm ct}=-\frac{1}{8 \pi G} \int_{\partial {\cal M}^{\pm 
}}d^{d-1}x\sqrt{-h}
\Biggl[
-\frac{d-2}{\ell}+\frac{\ell \Theta(d-4)}{2(d-3)}{\cal{R}}
-\frac{\ell^3 \Theta(d-6)}{2(d-5)(d-3)^2}
\Big({\cal R}_{AB}{\cal R}^{AB}-\frac{d-1}{4(d-2)}{\cal R}^2 \Big)
\Bigg]\ .
\end{eqnarray}
Here $ {\cal R}_{AB},~{\cal R}$ are the Ricci 
tensor and the Ricci scalar for the boundary metric and 
$\Theta(x)$ 
is the step function, which is equal to 1 for $x\geq 0$ and zero 
otherwise; 
$A,B,\ldots $ indicate the intrinsic coordinates of the boundary.
 $\int_{\mathcal{\partial M}^{\pm }}$ indicates the sum of the
integral over the early and late time boundaries. 
In what follows, to simplify the picture,
 we will consider the  $\mathcal{I}^{+}$ boundary only,
dropping the $^{\pm }$ indices (similar results hold for 
$\mathcal{I}^{-}$).

Using $I=I_{bulk}+I_{surf}+I_{ct}$,  one can
construct a  boundary stress tensor, which is given by 
the 
variation of the total action at the boundary with respect to 
$h_{AB}$ 
(its explicit expression is given in Ref.
\cite{Ghezelbash:2001vs}).

The next step is to write the boundary metric
 in a ADM-like general form 
\begin{equation}
ds^{  2}=h_{AB}^{  }d {x}^{A}d {x}^{B}=N_{t}^{ 2}dt ^{2}
+\sigma _{ab}^{ }\left( d\psi ^{  a}+N^{  a}dt
\right) \left( d\psi ^{ b}+N^{  b}dt \right) ,  \label{hmetric}
\end{equation}%
where $N_{t }$ and $N^{a}$ are the lapse function and the shift 
vector
respectively and the $\psi ^{a}$ are the intrinsic coordinates on the 
closed
surfaces $\Sigma $ (a $d-2$ dimensional sphere in our case).  
In this
approach, the conserved  quantity associated with a Killing vector 
$\xi
^{ i }$ on the $\mathcal{I}^{+}$ boundary  is  given by 
\begin{equation}
\label{Qcons}
\mathfrak{Q_{\xi}}{}^{  }=\oint_{\Sigma ^{ }}d^{d-2}\psi ^{  
}\sqrt{%
\sigma ^{  }}n^{A}T_{AB}^{  }\xi ^{B} ,  
\end{equation}%
where $n^{A}$ is an outward-pointing unit vector, normal to surfaces 
of
constant $r$.  Physically, this means that a collection of
observers, on the hypersurface with the induced metric $h_{AB}$, 
would all
measure the same value of $\mathfrak{Q}_{\xi }$ provided this surface 
has an
isometry generated by $\xi ^{i}$. 
If $\partial /\partial t$ is a Killing vector on $\Sigma$, then 
the conserved mass is defined to be the conserved quantity 
$\mathfrak{M}$ associated with it.

We have applied this approach to compute at the far future boundary
(outside the cosmological horizon) the mass of the solutions 
of the  $F(2)+F(4)$ model. The crucial point here is that these 
solutions approaches
asymptotically a Schwarzschild-dS background, the YM manifesting 
only
in the next to leading order of the $T_{AB}$ expression. As a result, 
one finds
\begin{eqnarray}
\label{Mct}
\mathfrak{M}=-\frac{(d-2)\Omega_{d-2}}{8 \pi G}M+E_0(d),
\end{eqnarray}
where $M$ is asymptotic value of the mass function $m(r)$
and $\Omega_{d-2}=2 \pi^{(d-1)/2}/\Gamma((d-1)/2)$ is the 
area of a unit $(d-2)$-dimensional sphere.
The additional term $E_0(d)$ appearing in (\ref{Mct}) for $d=5,(7)$
is the mass of pure global  dS$_{5,(7)}$ spacetime and is usually interpreted 
as the 
energy dual to the Casimir energy  of the CFT defined on a four 
(six)
dimensional Euclidean Einstein universe \cite{Ghezelbash:2001vs}
(i.e.  $E_0=3\Omega_3\ell^2/(64\pi G)$ for $d=5$ and 
$E_0=5\Omega_5\ell^4/(128\pi G)$ in the seven dimensional case).
One should also remark that all solutions of the  $F(2)+F(4)$ model 
we found 
have $M>0$ (both black holes and particle like solutions).
Thus  $\mathfrak{M}-E_0(d)$ is negative, consistent with the
expectation \cite{Balasubramanian:2001nb} that pure dS spacetime has 
the
largest mass for a singularity-free spacetime. 

However, one can easily see that this approach fails to assign a 
finite mass to  the 
solution of the EYM-$F(2)$ model, despite their asymptotically dS 
behaviour. (Note that this prescription regularizes the mass and action 
of the embedded abelian solutions \cite{Astefanesei:2003gw}). 
Asymptotically AdS solutions with a diverging ADM mass have been 
considered by
many authors, mainly for a scalar field in the bulk
(see e.g. \cite{Hollands:2005wt}). In this case it 
might be
possible to relax the standard asymptotic
conditions without loosing the original symmetries,
but modifying the charges in order to take into account the presence 
of matter fields.
For $\Lambda<0$, the Ref. \cite{Radu:2005mj} suggested 
that
it is still  possible to obtain a finite mass of EYM solutions in a 
$F(2)$ theory by
allowing $I_{ct}$ to depend not only on the boundary metric $h 
_{AB}$, but
also on the gauge field strength tensor. This means that the
quasilocal stress-energy tensor  also
acquires a contribution coming from the matter fields.

A similar approach holds also for dS solutions
in $d>4$ dimensions and 
we find that by adding
to the expression (\ref{ct}) a supplementary 
matter counterterm of the form
\begin{eqnarray}
I_{\mathrm{ct}}^{(m)} &=&-\frac{\tau_1}{4}\int_{\partial M} 
d^{4}x\sqrt{h}
\log (\frac{r}{\ell})~{\rm tr}\{ F_{AB}F^{AB}\},
\end{eqnarray}%
for $d=5$ and 
\begin{eqnarray}
\label{Ict-mat}
I_{ct}^{(m) }=
-\frac{\tau_1}{(d-5)}
 \int_{\partial M}d^{d-1}x\sqrt{h }
 ~{\rm tr}\{ F_{AB}F^{AB}\}, 
\end{eqnarray}
for $d>5$,
the mass divergence disappears.
This yields a supplementary contribution to the boundary 
stress-tensor
\begin{eqnarray}
\label{TAB-mat}
\nonumber
T_{AB}^{(m)}=-\frac{\tau_1\log (r/\ell)}{32\pi G} h_{AB}{\rm tr}\{F_{CD}F^{CD}\},
~~{\rm if}~d=5,~{\rm and}~~ 
T_{AB}^{(m)}=-\frac{1}{8\pi G}\frac{1}{d-5}h_{AB}~{\rm 
tr}\{F_{CD}F^{CD}\},
~~{\rm for}~d>5.
\end{eqnarray}
The mass of the $d>5$ solutions computed in this way 
is finite
\begin{eqnarray}
\label{mass-ctm}
\mathfrak{M}=-\frac{(d-2)\Omega_{d-2}}{8 \pi G} M_0+E_0(d), 
\end{eqnarray}
where $M_0$ is the constant appearing in the asymptotic
expansion (\ref{mass-div2}).
 
\section{Conclusions}
This work was primarily motivated by the question of how a positive
cosmological constant will affect the properties of the gravitating 
nonabelian
field solutions in a higher dimensional spacetime.
To the best of our knowledge, this question has not yet been 
addressed in the literature.
Apart from this motivation, the study of gravitating matter field 
configurations
in asymptotically dS space may help a better understanding of the 
conjectured dS/CFT
correspondence as well as act as a probe for the brane-world 
scenario.

Our findings have completed our qualitative understanding of 
gravitating nonabelian
solutions in higher ($d\ge 5$) dimensions, encompassing all possible 
values of the
cosmological constant $\Lambda<0$, $\Lambda=0$ and $\Lambda>0$, the 
last being the
object of the present investigation. Certain features of
these solutions are shared, while others differ. At the most basic 
level, we have seen
to varying degrees of rigour, that spherically symmetric
solutions to the usual ($p=1$), $F(2)$ YM model in
all these three cases have infinite mass in higher ($d\ge 5$) 
dimensions. By contrast, finite
masses are obtained when the YM sector of the model is augmented by 
the appropriate higher order ($p\ge 2$), $F(2p)$ members of the YM
hierarchy in all three cases.
(We have also considered the alternative option of using the 
counterterm method, avoiding the use of $p\ge 2$, $F(2p)$ terms.)
It can therefore be stated that, to
construct finite mass solutions  of gravitating nonabelian matter, 
the YM sector of the
theory must be an appropriate superposition of members of the YM 
hierarchy, beyond the
usual YM term. This statement can be qualified in the present 
context, namely when
$\Lambda>0$, by adding that even when the total mass diverges the 
mass {\it inside} the cosmological horizon is finite.
In addition to this salient feature of higher dimensional EYM 
solutions, our results reveal detailed qualitative properties
occurring in all three cases.


In the context of the $p=1$, $F(2)$ YM model with $\Lambda>0$
in $d\ge 5$ dimensional
spacetime, the most remarkable property is that the asymptotics of the
solutions lead to monopole-like configurations with nonvanishing 
magnetic flux. Indeed in those cases the solutions {\it never} have
instanton like asymptotics. This
circumstance makes it very easy to conclude that the total mass of 
these solutions is divergent.

In the context of the $F(2)+F(4)$ model studied in this paper,
which is the simplest case of models with
superposed $F(2p)$ terms, our results have led to an overview of the 
main qualitative features common to EYM solutions in all three cases
($\Lambda<0,\ \Lambda=0,\ \Lambda>0$), apart from their masses being 
finite. These confirm and expand on our knowledge of the branch patterns
of these solutions in various dimensions, learnt from the results of 
\cite{Brihaye:2002hr,Brihaye:2002jg} for $\Lambda=0$, and
\cite{Radu:2005mj} for $\Lambda<0$. The results of $\Lambda=0$
solutions were analytically analysed in \cite{Breitenlohner:2005hx} 
leading to a patterns that repeat modulo $d=4p+1$ dimensions. In
particular there arise two types of
patterns, those in dimensions $d=4p+1$ and the rest in $4p+2\le d\le 
4p+4$. In the restricted
context here, these are the dimensions $d=5$ on the one hand, and 
$6\le d\le 8$ on the other.
We have learnt here that these patterns arise also for the 
$\Lambda>0$ case. In particular
for the second case, solutions in $d=8$ conform to the pattern 
displayed on Figure 6b in
both the other cases. Much more interestingly the situation in $d=5$, 
which is qualitatively the same for $\Lambda<0$ and $\Lambda=0$ models,
strongly departs for $\Lambda>0$, from the patterns of the former. These
features are displayed on Figure 6a, and described in Section 3.2.

  What has not been studied quantitatively here, and in the
$\Lambda=0$~\cite{Brihaye:2002hr,Brihaye:2002jg,Breitenlohner:2005hx} and
$\Lambda<0$~\cite{Radu:2005mj} cases is the question of the stability
of the solutions. We expect that in all even spacetime dimensions, as well
as all odd dimensions $d=4p+1$, the solutions will be sphalerons like
the four dimensional Bartnik-McKinnon solutions~\cite{Bartnik:1988am}.
This is because in all these cases there is no topologically stable
soliton in the gravity decoupling limit. In all other odd spacetime
dimensions however, a stable soliton will survaive in the
gravity decoupling limit, stabilised by the $\frac{d-1}{2}$-th
Chern-Pontryagin (CP) charge, rather like the monopole charge in the
case of the gravitating monopole~\cite{Breitenlohner:1994di},
stabilised by the monopole charge (that descends from the $2$-nd
CP charge).

 As an additional remark, we allude to the case when the coupling
strength of the $F(4)$ term, $\tau_2$, takes on a negative value. In the
absence of existence proofs for solutions to models with higher order
YM terms, one cannot {\it a priori} exclude this possibility, especially
in view of the discovery of such explicit solutions in the usual
Einstein--Maxwell theory~\cite{Gibbons:1996pd}.
We have tried to construct such
solutions numerically and failed for models with $\Lambda\le 0$, but
surprisingly find them for the $\Lambda>0$ case at hand. 

While the present work is concerned with higher dimensional EYM solutions
with positive cosmological constant, we have at every stage  compared
our results to the corresponding ones pertaining to the asymptotically
flat~\cite{Brihaye:2002hr,Brihaye:2002jg} and the asymptotically
AdS~\cite{Radu:2005mj} counterparts. The general characteristics of both
$\Lambda\le 0$ solutions, at least of the important finite mass ones,
are quantitatively similar, while the corresponding characteristics of
$\Lambda>0$ case differ from the former strikingly, in a systematic
way. The numerically discovered features for $\Lambda=0$ solutions
are explained analytically in \cite{Breitenlohner:2005hx}, which can be
adapted systematically to the $\Lambda<0$ case which is qualitatively
similar. But an analytic study like \cite{Breitenlohner:2005hx} for the
$\Lambda>0$ case, using the methods of \cite{Breitenlohner:1993es} 
is outstanding and
is desirable to complete the overall comparative study of all three types
($\Lambda>0, \Lambda=0$ and $\Lambda<0$) of higher dimensional
EYM solutions.
\\
\\
\\
\\
{\bf Acknowledgement}
\\
We are grateful to the referee for valuable remarks which helped us to improve our manuscript.
This work was carried out in the framework of Enterprise--Ireland
Basic Science Research Project SC/2003/390.
YB is grateful to the Belgian FNRS for financial support.

 \newpage



\begin{thebibliography}{99}
\bibitem{Tangherlini:1963bw}
  F.~R.~Tangherlini,
  Nuovo Cim.\  {\bf 27} (1963) 636.
\bibitem{Myers:1986un}
  R.~C.~Myers and M.~J.~Perry,
  Annals Phys.\  {\bf 172} (1986) 304.
\bibitem{Volkov:2001tb}
  M.~S.~Volkov,
  Phys.\ Lett.\ B {\bf 524} (2002) 369
  [arXiv:hep-th/0103038].
\bibitem{Brihaye:2002hr}
  Y.~Brihaye, A.~Chakrabarti and D.~H.~Tchrakian,
  Class.\ Quant.\ Grav.\  {\bf 20} (2003) 2765
  [arXiv:hep-th/0202141].
\bibitem{Brihaye:2002jg}
  Y.~Brihaye, A.~Chakrabarti, B.~Hartmann and D.~H.~Tchrakian,
  Phys.\ Lett.\ B {\bf 561} (2003) 161
  [arXiv:hep-th/0212288].
\bibitem{Bartnik:1988am}
  R.~Bartnik and J.~McKinnon,
  Phys.\ Rev.\ Lett.\  {\bf 61} (1988) 141.
\bibitem{Breitenlohner:1992}
P.~Breitenlohner, P.~Forgacs and D.~Maison, Nucl. Phys.
{\bf B 383} (1992) 357; {\it ibid.} {\bf 442} (1995) 126
\bibitem{Breitenlohner:2005hx}
  P.~Breitenlohner, D.~Maison and D.~H.~Tchrakian,
  Class.\ Quant.\ Grav.\  {\bf 22} (2005) 5201
  [arXiv:gr-qc/0508027].
\bibitem{Okuyama:2002mh}
N.~Okuyama and K.~i.~Maeda,
Phys.\ Rev.\ D {\bf 67} (2003) 104012
[arXiv:gr-qc/0212022].
\bibitem{Radu:2005mj}
  E.~Radu and D.~H.~Tchrakian,
  Phys.\ Rev.\ D {\bf 73} (2006) 024006
  [arXiv:gr-qc/0508033].
\bibitem{Winstanley:1998sn}
E.~Winstanley,
Class.\ Quant.\ Grav.\  {\bf 16} (1999) 1963
[arXiv:gr-qc/9812064].
\bibitem{Bjoraker:2000qd}
J.~Bjoraker and Y.~Hosotani,
Phys.\ Rev.\ D {\bf 62} (2000) 043513
[arXiv:hep-th/0002098].
\bibitem{Volkov:1996qj}
  M.~S.~Volkov, N.~Straumann, G.~V.~Lavrelashvili, M.~Heusler and 
O.~Brodbeck,
  Phys.\ Rev.\ D {\bf 54} (1996) 7243
  [arXiv:hep-th/9605089].

\bibitem{Linden:2000ev}
  A.~N.~Linden,
  Commun.\ Math.\ Phys.\  {\bf 221} (2001) 525
  [arXiv:gr-qc/0005004].

\bibitem{Brihaye:2006kn}
  Y.~Brihaye, B.~Hartmann, E.~Radu and C.~Stelea,
  arXiv:gr-qc/0607078.

\bibitem{Torii:1995wv}
  T.~Torii, K.~i.~Maeda and T.~Tachizawa,
  Phys.\ Rev.\ D {\bf 52} (1995) 4272
  [arXiv:gr-qc/9506018].

\bibitem{Breitenlohner:2004fp}
  P.~Breitenlohner, P.~Forgacs and D.~Maison,
  Commun.\ Math.\ Phys.\  {\bf 261} (2006) 569
  [arXiv:gr-qc/0412067].
\bibitem{Gibbons:2006wd}
  G.~W.~Gibbons and P.~K.~Townsend,
  Class.\ Quant.\ Grav.\  {\bf 23} (2006) 4873
  [arXiv:hep-th/0604024].
 
\bibitem{COLSYS}
 U. Ascher, J. Christiansen, R.~D. Russell,
 Math. of Comp. {\bf 33} (1979) 659;\\
 U. Ascher, J. Christiansen, R.~D. Russell,
 ACM Trans. {\bf 7} (1981) 209.
\bibitem{Tseytlin}
A.A. Tseytlin, {\it Born--Infeld action, supersymmetry and string 
theory},
in {\it Yuri Golfand memorial volume}, ed. M. Shifman, World 
Scientific (2000).
\bibitem{BRS}
E. Bergshoeff, M. de Roo and A. Sevrin, Fortsch.Phys. 49 (2001) 
433-440;
Nucl.Phys.Proc.Suppl. 102 (2001) 50-55.
  E.~A.~Bergshoeff, M.~de Roo and A.~Sevrin,
  Fortsch.\ Phys.\  {\bf 49} (2001) 433
  [Nucl.\ Phys.\ Proc.\ Suppl.\  {\bf 102} (2001) 50]
  [arXiv:hep-th/0011264].
\bibitem{CNT}
  M.~Cederwall, B.~E.~W.~Nilsson and D.~Tsimpis,
  JHEP {\bf 0106} (2001) 034
  [arXiv:hep-th/0102009]. 
\bibitem{Balasubramanian:1999re}
V.~Balasubramanian and P.~Kraus,
Commun.\ Math.\ Phys.\  {\bf 208} (1999) 413
[arXiv:hep-th/9902121].
\bibitem{Ghezelbash:2001vs}
  A.~M.~Ghezelbash and R.~B.~Mann,
  JHEP {\bf 0201} (2002) 005
  [arXiv:hep-th/0111217].
\bibitem{Balasubramanian:2001nb}
  V.~Balasubramanian, J.~de Boer and D.~Minic,
  Phys.\ Rev.\ D {\bf 65} (2002) 123508
  [arXiv:hep-th/0110108].
\bibitem{Astefanesei:2003gw}
  D.~Astefanesei, R.~Mann and E.~Radu,
  JHEP {\bf 0401} (2004) 029
  [arXiv:hep-th/0310273].
\bibitem{Hollands:2005wt}
  S.~Hollands, A.~Ishibashi and D.~Marolf,
  Class.\ Quant.\ Grav.\  {\bf 22} (2005) 2881
  [arXiv:hep-th/0503045];
\\
  T.~Hertog and K.~Maeda,
  JHEP {\bf 0407} (2004) 051
  [arXiv:hep-th/0404261];
\\
  T.~Hertog and K.~Maeda,
  Phys.\ Rev.\ D {\bf 71} (2005) 024001
  [arXiv:hep-th/0409314];
\\
  M.~Henneaux, C.~Martinez, R.~Troncoso and J.~Zanelli,
  Phys.\ Rev.\ D {\bf 70} (2004) 044034
  [arXiv:hep-th/0404236];
\\
  E.~Radu and D.~H.~Tchrakian,
  Class.\ Quant.\ Grav.\  {\bf 22} (2005) 879
  [arXiv:hep-th/0410154];
\\
  J.~T.~Liu and W.~A.~Sabra,
  Phys.\ Rev.\ D {\bf 72} (2005) 064021
  [arXiv:hep-th/0405171].
\bibitem{Breitenlohner:1993es}
  P.~Breitenlohner, P.~Forgacs and D.~Maison,
  Commun.\ Math.\ Phys.\  {\bf 163} (1994) 141.
\bibitem{Gibbons:1996pd}
  G.~W.~Gibbons and D.~A.~Rasheed,
  Nucl.\ Phys.\ B {\bf 476} (1996) 515
  [arXiv:hep-th/9604177].
\bibitem{Breitenlohner:1994di}
  P.~Breitenlohner, P.~Forgacs and D.~Maison,
  Nucl.\ Phys.\ B {\bf 442} (1995) 126
  [arXiv:gr-qc/9412039].


 


 


 


 


 
\end{thebibliography}
\end{document}